\documentclass[aps,onecolumn,11pt,prc]{revtex4-2}
\usepackage{graphicx}
\usepackage{amssymb,amsmath,amscd,amsthm}
\usepackage{times}
\newcommand{\ba}{\begin{eqnarray}}
\newcommand{\ea}{\end{eqnarray}}
\newcommand{\bsub}{\begin{subequations}}
\newcommand{\esub}{\end{subequations}}
\newcommand\tb{\tilde{\beta}}

\def\ket#1{|#1\rangle}
\def\bra#1{\langle#1|}

\usepackage[percent]{overpic}  
\usepackage{braket}

\begin{document}

\sloppy \raggedbottom

\title{Linking partial dynamical symmetry to nuclear energy density functionals}

\author{A. Leviatan$^1$}{}, 
\author{N. Gavrielov$^1$}{},
\author{K. Nomura$^2$}{}

\affiliation{$^1$Racah Institute of Physics, The Hebrew University,
  Jerusalem 91904, Israel}{}
\affiliation{$^2$Department of Physics, Faculty of Science, University of
  Zagreb, HR-10000 Zagreb, Croatia}{}

\begin{abstract}
  We use self-consistent mean-field methods in combination with
  the interacting boson model (IBM) of nuclei, to establish
  a linkage between universal energy density functionals (EDFs)
  and partial dynamical symmetry (PDS). An application to $^{168}$Er
  shows that IBM Hamiltonians derived microscopically from known
  nonrelativistic and relativistic EDFs in this region, conform
  with SU(3)-PDS.
\end{abstract}
\maketitle

\section{Introduction}
The concept of dynamical symmetry (DS)
is by now widely recognized to play a key role in the
structure of nuclei.
Its basic paradigm is to write the Hamiltonian of the
system in terms 
of the Casimir operators of a chain of nested algebras, 
$G_{\rm dyn} \supset G_1 \supset G_2 \supset \dots \supset G_{\rm sym}$,
where $G_{\rm dyn}$ is the dynamical (spectrum generating) 
algebra of the system such that operators of all physical
observables can be written in terms of its generators
and $G_{\rm sym}$ is the symmetry algebra~\cite{LieAlg}.
A dynamical symmetry is characterized by complete solvability
for {\it all} states in terms of quantum numbers, which are
the labels of irreducible representations (irreps)
of the algebras in the chain.

A notable example of such algebraic construction
is the interacting boson model
(IBM)~\cite{IBM},
which describes low-lying quadrupole
collective states in nuclei in terms of $N$ monopole
$(s)$ and quadrupole $(d)$ bosons. In this case, 
$G_{\rm dyn}\!=\!{\rm U(6)}$ and $G_{\rm sym}\!=\!{\rm SO(3)}$.
The model accommodates several DS chains with leading
subalgebra
$G_1\!=\! {\rm U(5)}$, ${\rm SU(3)}$ and ${\rm SO(6)}$,
whose spectra resemble known paradigms of nuclear
collective structure: spherical vibrator,
axially-deformed rotor
and $\gamma$-soft deformed rotor, respectively.

Geometry is introduced in the IBM through an energy
surface,
\ba
E_{\rm IBM}(\tb,\gamma) &=&
\bra{\tb,\gamma; N} \hat{H} \ket{\tb,\gamma ; N} ~,
\label{enesurf}
\ea 
defined by the expectation value of the Hamiltonian in
a coherent state~\cite{gino80,diep80},
\bsub
\ba
\vert\tb,\gamma ; N \rangle &=&
(N!)^{-1/2}[b^{\dagger}_{c}(\tb,\gamma)]^N
\,\vert 0\,\rangle ~,
\\
b^{\dagger}_{c}(\tb,\gamma)
&=& (1+\tb^2)^{-1/2}[\tb\cos\gamma 
d^{\dagger}_{0} + \tb\sin{\gamma} 
(d^{\dagger}_{2} + d^{\dagger}_{-2})/\sqrt{2}
+ s^{\dagger}]
~.\qquad
\ea
\label{int-state}
\esub
Here $(\tb,\gamma)$ are
quadrupole shape parameters in the IBM,
whose values, $(\beta_0,\gamma_0)$,
at the global minimum of $E_{\rm IBM}(\tb,\gamma)$
define the equilibrium 
shape for a given Hamiltonian. 
The shape can be spherical $(\beta_0 \!=\!0)$ or 
deformed $(\beta_0 \!>\!0)$ with
$\gamma_0 \!=\!0^{\circ}$ (prolate), 
$\gamma_0 \!=\!60^{\circ}$ (oblate), 
$0^{\circ} \!<\! \gamma_0 \!<\! 60^{\circ}$ (triaxial) or
$\gamma$-independent.
The equilibrium deformations associated with the 
DS limits, conform with their geometric interpretation,
and are given by 
$\beta_{0}\!=\!0$ for U(5), 
$(\beta_0 \!=\!\sqrt{2},\gamma_{0}\!=\!0^{\circ})$ for SU(3)
and $(\beta_{\rm 0}\!=\!1,\gamma_{0}\,\,{\rm arbitrary})$
for SO(6). The coherent state
$\ket{\beta_0,\gamma_0;N}$~(\ref{int-state}),
with the equilibrium deformations, serves as the
intrinsic state for the ground~band.

The merits of a DS, with its analytic and geometric
attributes, are self evident. 
However, in the majority of nuclei,
an exact DS rarely occurs and one is compelled to break it.
More often some states obey the patterns required by the
symmetry, but others do not.
The need to address such situations, but still preserve
important symmetry remnants, has motivated
the introduction of partial dynamical symmetry
(PDS)~\cite{leviatan2011,leviatan1996}.
The essential idea is to relax
the stringent conditions imposed by an exact DS
so that solvability and/or good quantum numbers are
retained by only a subset of states.
Detailed studies in the IBM framework,
have shown that PDSs account quite well
for a wealth of spectroscopic data in various types of
nuclei~\cite{leviatan2011,leviatan1996,LevSin99,
  casten2014,leviatan2013,ramos2009,vanisacker2015}
and are relevant to related quantum phase transitions
and shape-coexistence~\cite{leviatan2007,macek2014,
leviatan2016,leviatan2017,leviatan2018}.
In all these phenomenological studies, an Hamiltonian
with a prescribed PDS is introduced, its parameters are
determined from a fit to the spectra, and the PDS
predictions (which are often parameter-free)
are compared with the available empirical energies
and transition rates.
In the present contribution, we show that
the PDS notion is robust and founded on
microscopic grounds~\cite{NomGavLev21}.

PDSs do not arise from invariance properties of the
Hamiltonian, hence can be referred to as emergent
symmetries. The role of an emergent Sp(3,R) DS
has been recently demonstrated
within ab-initio calculations of light
nuclei~\cite{dytrych2020,mccoy2020}.
Here we focus on heavy nuclei, and present an
efficient procedure to uncover the
microscopic origin of PDS by linking it to
universal nuclear energy density functionals.
We apply the~procedure to $^{168}$Er,
in which the SU(3)-PDS was
previously recognized on phenomenological
grounds~\cite{leviatan1996,LevSin99,casten2014}.

\section{SU(3) PDS: a phenomenological approach}
\label{sec:su3-ds}

The SU(3) dynamical symmetry limit and basis states
correspond to the chain,
\ba
   {\rm U(6)\supset SU(3)\supset SO(3)}
\qquad \ket{[N](\lambda,\mu)KL} ~.
\label{su3-chain}
\ea   
The SU(3)-DS Hamiltonian involves the Casimir operators,
$\hat{C}_{\rm G}$, of the algebras in the chain.
The spectrum consists of SU(3)
multiplets with the states $\ket{[N](\lambda,\mu)KL}$
specified by the total boson number $N$, 
the SU(3) irrep $(\lambda,\mu)$, the angular
momentum $L$, and the label $K$ which
corresponds to the projection of the 
angular momentum on the symmetry axis. 
The lowest multiplets have
$(\lambda,\mu)\!=\!(2N,0)$ which contains
the ground band~$g(K\!=\!0)$,
and $(\lambda,\mu)\!=\!(2N-4,2)$
which contains both
the $\beta(K\!=\!0)$ and $\gamma(K\!=\!2)$ bands.

Following the general
algorithm~\cite{leviatan2011,ramos2009,AL92},
the SU(3)-PDS Hamiltonian is constructed to
be~\cite{leviatan1996},
\ba
\hat{H}_{\rm PDS} = h_{0} P^{\dag}_0P_{0}
+ h_2 P^{\dag}_2\cdot\tilde{P}_{2}
+ \rho\,\hat{L}\cdot\hat{L} ~.
\label{h-PDS}
\ea
Here $P^{\dag}_0 = d^{\dag}\cdot d^{\dag}
- 2(s^{\dag})^2$,
$P^{\dag}_{2m} =
2d^{\dag}_{m}s^{\dag}
+ \sqrt{7}(d^{\dag}d^{\dag})^{(2)}_{m}$,
$\tilde{P}_{2m}=(-1)^{m}{P_{2,-m}}$,
$\hat{L}$ the angular momentum operator and
standard notation of angular momentum coupling is used.
For $h_0=h_2$, $\hat{H}_{\rm PDS}$ reduces to the
SU(3)-DS Hamiltonian $\hat{H}_{\rm DS} = h_2[
  -\hat{C}_{\rm SU(3)} + 2\hat{N}(2\hat{N}+3)]
+ \rho\,\hat{C}_{\rm SO(3)}$.
For $h_0\neq h_2$, $\hat{H}_{\rm PDS}$ is no longer diagonal
in the SU(3)-DS chain~(\ref{su3-chain}),
but still has a subset of
eigenstates with good SU(3) symmetry. This comes about
because $P_0$ and $P_{2m}$ annihilate the states
$\ket{[N](2N,0)K\!=\!0,L}$ comprising the ground band
$g(K\!=\!0)$ and $P_0$ annihilates also the states
$\ket{[N](2N-4k,2k)K\!=\!2k,L}$
comprising the $\gamma^{k}(K\!=\!2k)$ bands.
In particular, the ground and gamma bands remain solvable with
good SU(3) quantum numbers, $(\lambda,\mu)=(2N,0)$ and
$(2N-4,2)$, and energies
\bsub
\ba
g(K\!=\!0):&& \qquad E =\rho L(L+1) ~,\\
\gamma(K\!=\!2):&&\qquad
E = h_{2}\,6(2N -1) + \rho L(L+1) ~,
\ea
\label{ene-PDS}
\esub
while the $\beta(K\!=\!0)$ band is mixed.
\begin{figure}[t]
\centerline{
\fbox{\includegraphics[width=0.52\linewidth]{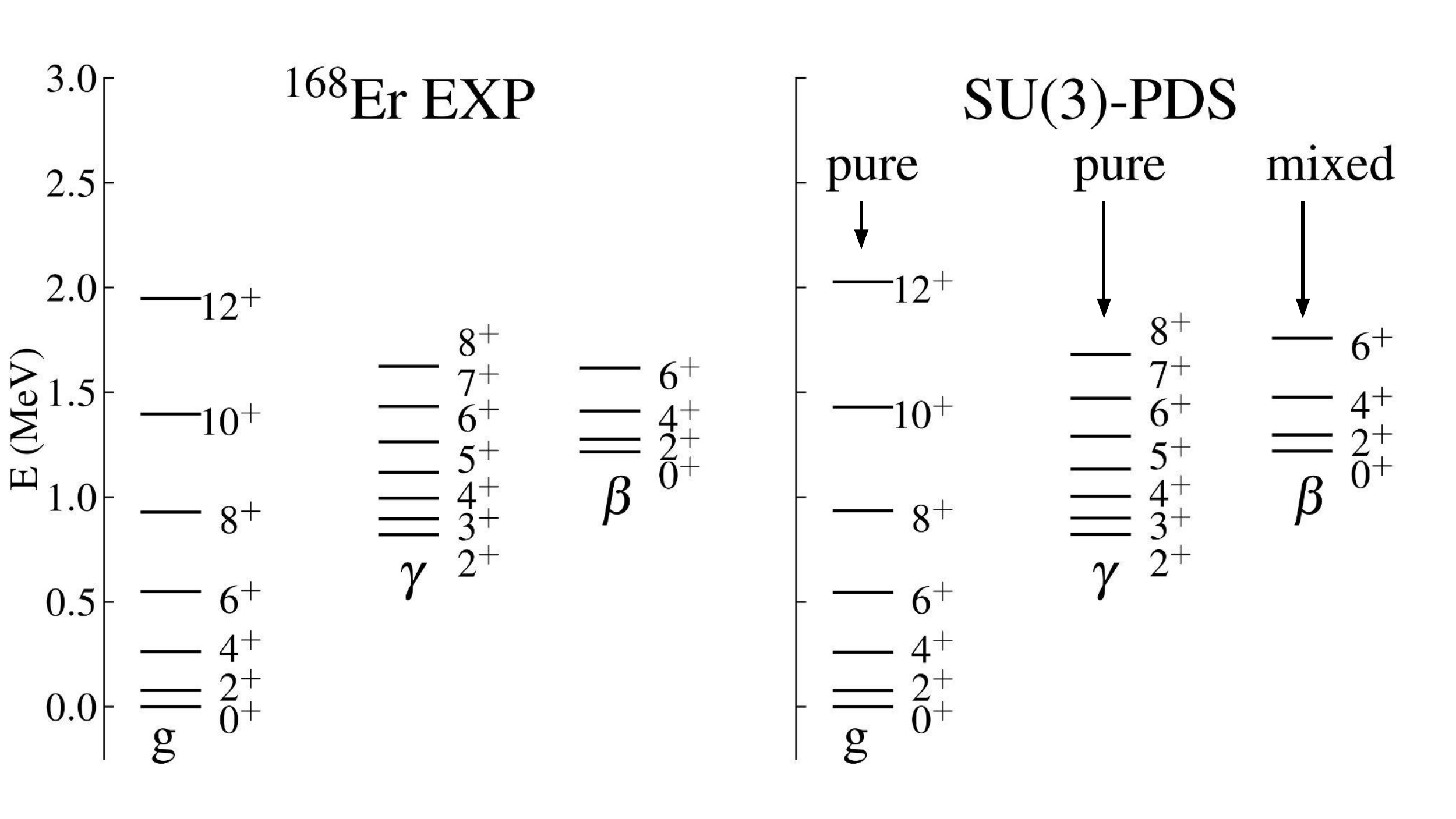}}
\hspace{0.06cm}%
\includegraphics[width=0.45\linewidth]{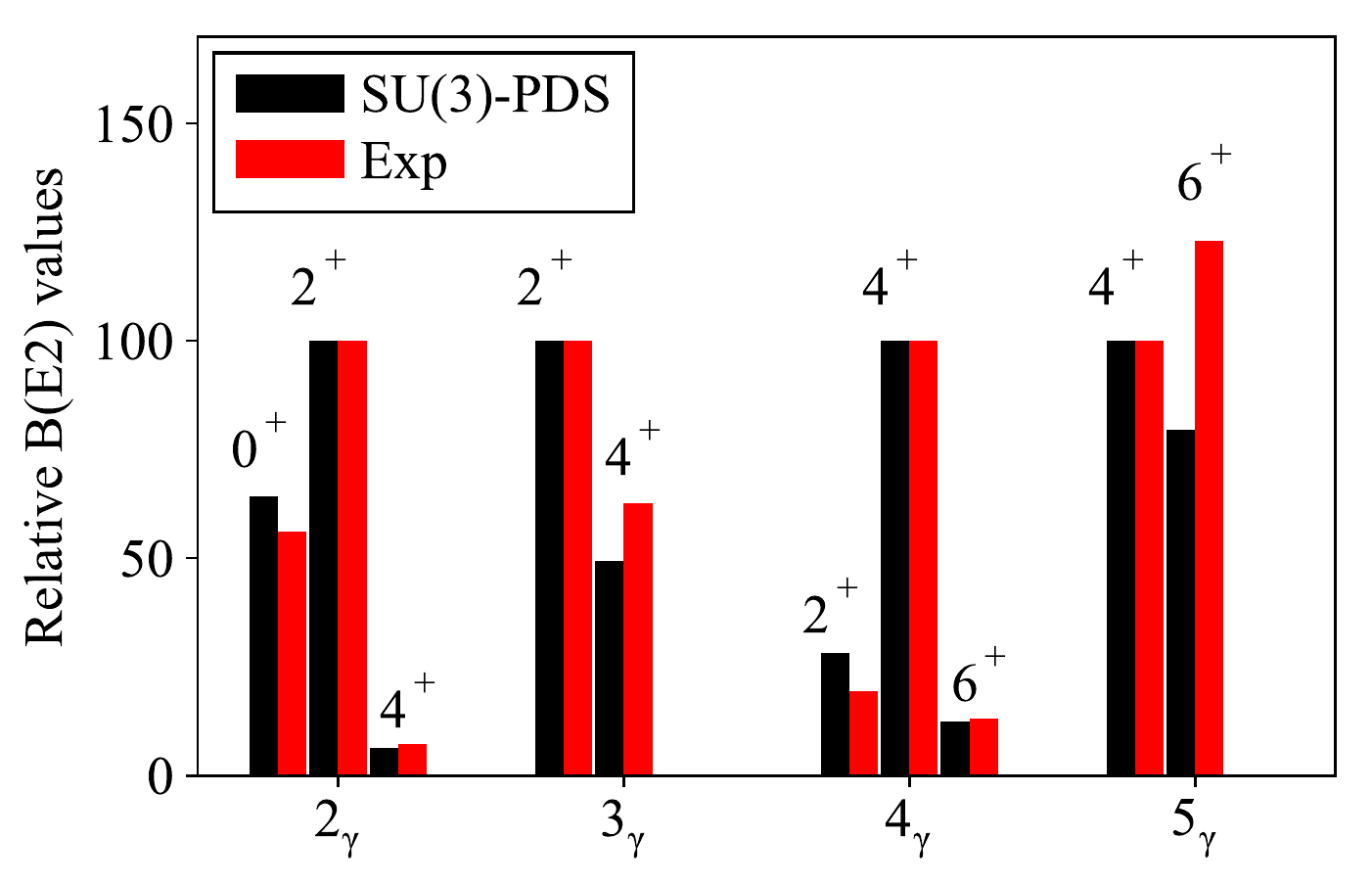}}
  \caption{
\small
Left panels: observed spectrum of $^{168}$Er compared to
an SU(3)-PDS calculation employing
$\hat{H}_{\rm PDS}$
of Eq.~(\ref{h-PDS})
with $h_0\!=\!8,\,h_2\!=\!4,\,\rho\!=\!13$ keV
and $N\!=\!16$, for which the ground and $\gamma$ bands
are pure while the $\beta$ band is mixed,
with respect to SU(3). Right panel:
comparison of the PDS parameter-free predictions with
the data on relative
$B(E2; L_{\gamma}\to L)$ values for $\gamma\to g$ $E2$
transitions in $^{168}$Er.
Adapted from ~\cite{leviatan1996,casten2014}.
\label{figEr168}}
\end{figure}

In a phenomenological approach, the parameters of
$\hat{H}_{\rm PDS}$ are determined from a fit to experimental
energies; $h_2$ and $\rho$ from $E(2_1)$ and $E(2_2)$,
using Eq.~(\ref{ene-PDS}),
and $h_0$ from $E(0_2)$.
As shown in Fig.~\ref{figEr168}, a~PDS calculation with
parameters indicated in the caption,
provides a good description for the
lowest bands in $^{168}$Er.~\cite{leviatan1996}.
The ground and gamma are pure SU(3) bands, but the beta
band is found to contain $13\%$ admixtures into the dominant
$(2N-4,2)$ irrep~\cite{LevSin99}.
Since the wave functions of the solvable states are known,
it is possible to obtain {\it analytic} expressions for
matrix elements of observables between them.
The $E2$ operator can be transcribed as
$\hat{T}(E2) =
\alpha\, Q^{(2)} + \theta\, (d^{\dag}s + s^{\dag}\tilde{d})$,
with $Q^{(2)}= d^{\dag}s + s^{\dag}\tilde{d}
-\tfrac{\sqrt{7}}{2}(d^{\dag}\tilde{d})^{(2)}$,
an SU(3) generator.
Since the solvable ground and gamma bands reside in
different SU(3) irreps, $Q^{(2)}$ cannot connect them and,
consequently, $B(E2)$ ratios for $\gamma\to g$ transitions
do no depend on the $E2$ parameters ($\alpha,\theta$)
nor on parameters of the PDS Hamiltonian~(\ref{h-PDS}).
Overall, as shown in the right panel of Fig.~1, these
parameter-free predictions of SU(3)-PDS
account well for the data in
$^{168}$Er~\cite{leviatan1996,casten2014}.

$\hat{H}_{\rm PDS}$ in Eq.~(\ref{h-PDS}) decomposes
naturally
into intrinsic and collective parts.
The former, consisting of the $h_0$ and $h_2$ terms,
determines the energy surface~(\ref{enesurf})
and band-structure, while the latter, consisting of the
$\rho$ term, determines the in-band rotational splitting.
Such a resolution is valid also for the general
IBM Hamiltonian describing the dynamics of
a prolate-deformed shape, which reads~\cite{leviatan1987},
\ba
\label{eq:ham}
\hat{H} = h_{0} P^{\dag}_0(\beta_0)P_{0}(\beta_0) + h_2
P^{\dag}_2(\beta_{0})\cdot\tilde{P}_{2}(\beta_0)
+ \rho\hat{L}\cdot\hat{L} ~.
\ea
Here $P^{\dag}_0(\beta_0) \!=\! d^{\dag}\cdot d^{\dag}
- \beta_{0}^{2}(s^{\dag})^2$ and 
$P^{\dag}_{2m}(\beta_0) =
\beta_0\sqrt{2}d^{\dag}_{m}s^{\dag}
+ \sqrt{7}(d^{\dag}d^{\dag})^{(2)}_{m}$.
Its energy surface, obtained from Eq.~(\ref{enesurf}),
is given by
\ba
\label{eq:pes}
E_{\mathrm{IBM}}(\tb,\gamma)
= N(N-1)(1+\tb^2)^{-2}[\,h_0(\tb^2-\beta^2_0)^2
  +2h_2\tb^2(\tb^2-2\beta_0\tb\cos{3\gamma}+\beta_0^2)\,] ~.
\ea
For $h_0,h_2\!\geqslant\! 0$, the surface has a
global minimum at
$(\tb\!=\!\beta_{0}\!>\!0,\gamma\!=\!0^{\circ})$,
corresponding to a prolate-deformed equilibrium shape.
The contribution of the rotational $\rho$-term
to the energy surface is $1/N$ suppressed, hence negligible.
$P_0(\beta_0)$ and $P_{2m}(\beta_0)$
annihilate the states with angular momentum $L$ projected
from the intrinsic state
$\ket{\beta_0,\gamma_0\!=\!0;N}$~(\ref{int-state}).
The Hamiltonian $\hat{H}$ of Eq.~(\ref{eq:ham}),
reduces to $\hat{H}_{\rm PDS}$ of Eq.~(\ref{h-PDS}), when
the following conditions are met,
\ba
   {\rm SU(3)\,PDS}:\quad h_0\neq h_2\;\;, \;\;
   \beta_0=\sqrt{2} ~.
\label{PDS-cond}
\ea   
In what follows, we show that IBM Hamiltonians
derived from microscopic considerations for $^{168}$Er, 
exhibit spectral properties of SU(3)-PDS.

\section{SCMF to IBM mapping}
The nuclear energy density functional (EDF) framework
allows for a reliable quantitative prediction
of ground-state properties and collective excitations of
nuclei over the
entire region of the nuclear chart.
Its basic implementation is in self-consistent
mean-field (SCMF) methods, in which an EDF is
constructed as a
functional of one-body nucleon density matrices that
correspond to a single product state. Pairing correlations
are taken into account by a choice of pairing force.
In the present contribution, we consider both 
nonrelativistic~\cite{bender2003,robledo2019} and
relativistic~\cite{vretenar2005,niksic2011} EDFs,
so as to ensure the robustness of the results.

The starting point is a set of constrained SCMF
calculations of an energy surface~\cite{RS}.
The constraints refer to those for mass
quadrupole moments, which are associated with the
deformation parameters $\beta$ and $\gamma$~\cite{BM_II}.
For the nonrelativistic SCMF calculations, we employ the
the Hartree-Fock plus BCS model~\cite{ev8,ev8r} with two
parameterizations of the Skyrme EDF~\cite{RS}
and a density-dependent delta force with strength $V_0$.
Specifically, the SLy4~\cite{SLy4} parameterization with
pairing strengths $V_{0}\!=\!1000$ and $1250$ MeVfm$^{3}$,
and the SkP~\cite{SkP} parameterization with $V_{0}\!=\!800$
and $1000$ MeVfm$^{3}$.
A smooth cut-off of 5 MeV below and above the Fermi surface
is invoked for these zero-range pairing forces~\cite{SkP}.
For the relativistic SCMF calculations, we employ the
relativistic Hartree-Bogoliubov
model~\cite{vretenar2005,DIRHB} with two types of EDFs.
Specifically, the density-dependent point-coupling (DD-PC1)~\cite{DDPC1}
and meson-exchange (DD-ME2)~\cite{DDME2} functionals,
both with a separable pairing force of finite range~\cite{tian2009}
and strengths $V_{0}\!=\!728$ and $837$ MeVfm$^{3}$, resembling
a finite-range Gogny interaction D1S.

\begin{figure*}[t]
\begin{center}
  \begin{tabular}{cc}
\includegraphics[width=0.46\linewidth]{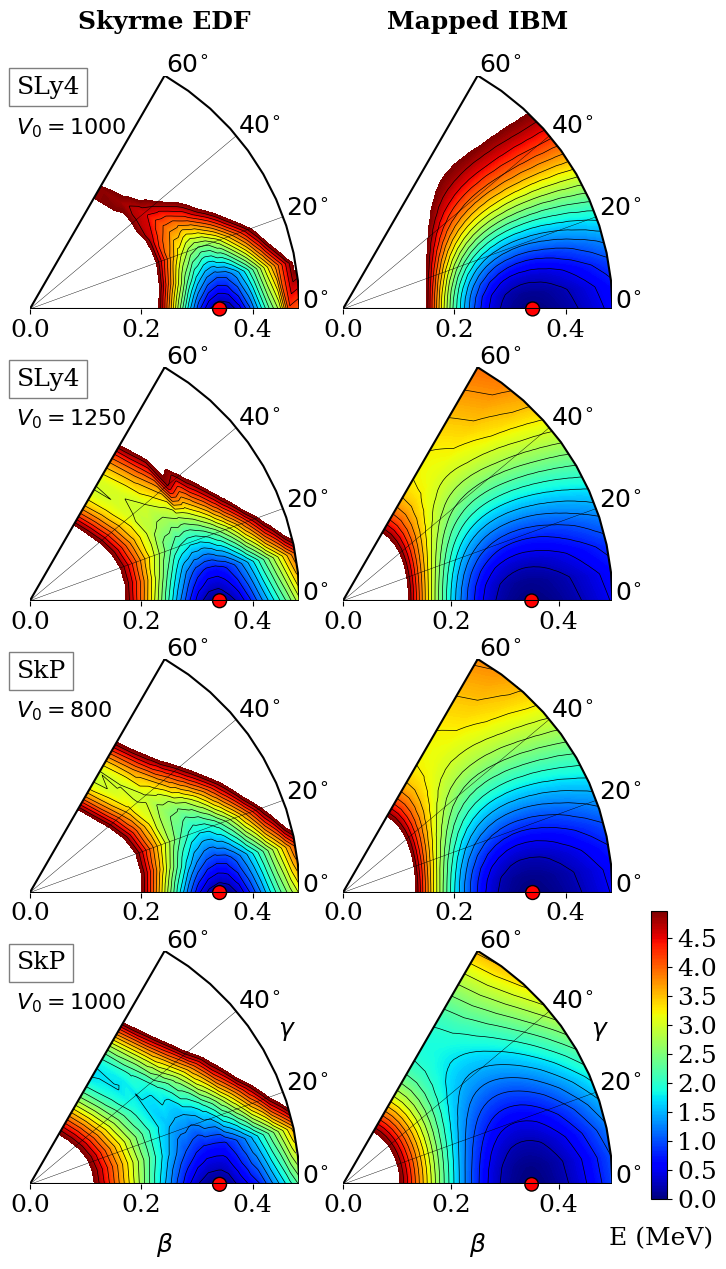} &
\includegraphics[width=0.46\linewidth]{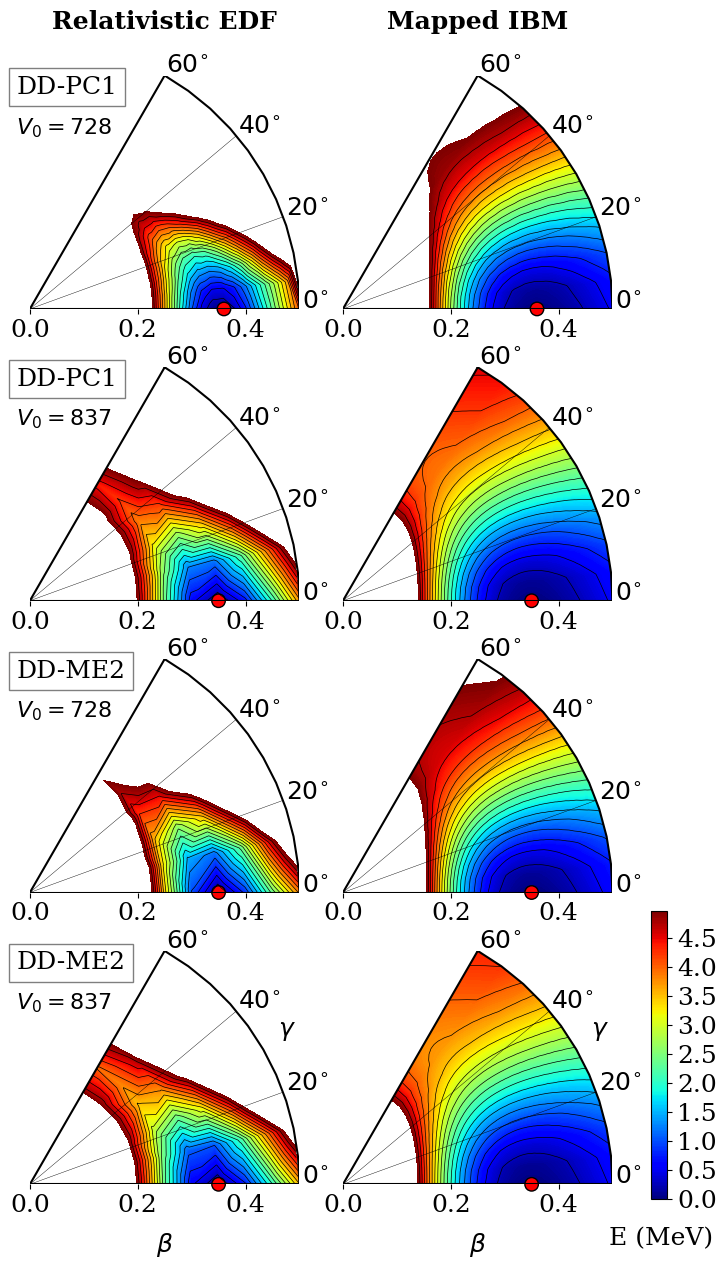} \\
\end{tabular}
\caption{SCMF energy surfaces in the $\beta$-$\gamma$
  plane for $^{168}$Er, based on
  the nonrelativistic Skyrme SLy4 and SkP EDFs
  (first column) and
 the relativistic DD-PC1 and DD-ME2 EDFs (third column)
 with different
 values of pairing strengths $V_{0}$ in units of MeVfm$^{3}$.
 The corresponding mapped IBM energy surfaces 
 are plotted on the second and fourth columns.
Contour spacing is 0.25 MeV, and 
the global minimum is indicated by a solid circle.
Adapted from~\cite{NomGavLev21}.
}
\label{fig:pes} 
\end{center}
\end{figure*}

The calculated SCMF energy surfaces $E_{\mathrm{SCMF}}(\beta,\gamma)$
for $^{168}$Er, based on the above nonrelativistic and relativistic EDFs,
are displayed on the first and third columns of Fig.~\ref{fig:pes},
respectively. As seen, all adopted  EDFs
lead to energy surfaces accommodating a pronounced
prolate-deformed
global minimum at $(\beta\!\approx\! 0.35,\gamma\!=\!0^{\circ})$.
The minimum tends to be less steep, in both the
$\beta$ and $\gamma$ directions,
for larger pairing strengths.
This is anticipated since
pairing correlations favor a more spherical shape.

From the ensemble of Hamiltonians
given in Eq.~(\ref{eq:ham}), the IBM Hamiltonian
appropriate for $^{168}$Er
is derived by the procedure developed
in~\cite{nomura2008,nomura2010,nomura2011rot}.
The parameters $\{h_{0}$, $h_{2}$, $\beta_{0}\}$
are determined by mapping the microscopic energy
surface $E_{\mathrm{SCMF}}(\beta,\gamma)$, obtained
for a given EDF, onto the corresponding IBM surface
$E_{\mathrm{IBM}}(\beta,\gamma)$ of Eq.~(\ref{eq:pes}).
The condition,
\ba
E_{\mathrm{SCMF}}(\beta,\gamma)\approx
E_{\mathrm{IBM}}(\beta,\gamma) ~,
\label{mapping}
\ea
is imposed to ensure
similar topology
in the neighborhood of the global minimum.
(The two surfaces are expressed in terms of $\beta$,
since the IBM and SCMF deformations are related by
$\tb=C\beta$, where the constant $C$ is determined by
the mapping). $N$ is fixed by
the usual boson counting, from the number
of valence nucleon pairs counted from the nearest
closed shell. The parameter $\rho$, Eq.~(\ref{eq:ham}),
is obtained by equating the
cranking moment of inertia in the IBM
to the Thouless-Valatin value \cite{delaroche2010},
the procedure discussed in detail
in~\cite{nomura2011rot}.
The mapped IBM energy surfaces, based on the
nonrelativistic and relativistic EDFs, are
shown on the second and fourth columns of
Fig.~\ref{fig:pes}, respectively.
One clearly sees that the IBM and microscopic surfaces
share common essential features near and
up to a few MeV above the global minimum.
In what follows, we examine to what extent the derived
EDF-based IBM Hamiltonians fulfill the
conditions~(\ref{PDS-cond}) for SU(3)-PDS.
\begin{table}[t]
  \caption[]{\label{tab:para}
 Parameters $h_{0},\,h_{2},\,\rho$ (in keV) and
$\beta_{0}$, of the Hamiltonian~(\ref{eq:ham})
 obtained from SCMF calculations based on
 nonrelativistic Skyrme SLy4 and SkP EDFs, and
 relativistic DD-PC1 and DD-ME2 EDFs, with
 pairing strengths $V_{0}$ (in MeV fm$^{3}$).
 The corresponding parameters for
 SU(3)-PDS~\cite{leviatan1996},
 are also shown. $E(2_2)$ and $E(0_2)$
 are the calculated
 bandhead energies (in keV) for the $\gamma$ and $\beta$ bands
 and $R=\tfrac{E(0_2)}{E(2_2)}$. For $^{168}$Er,
$E(2_2)\!=\!821,\,E(0_2)\!=\!1217$ (in keV)
 and $R\!=\!1.48$~\cite{nndc}.
 Adapted from~\cite{NomGavLev21}.}\small\smallskip
\begin{center}

  \tabcolsep=3.6pt
  \begin{tabular}{lccccc|ccc}
\hline
\hline
&&&&&&&&\\[-8pt]
    EDF & $V_{0}$  & $h_0$ & $h_2$ & $\rho$ & $\beta_0$ &
    $E(2_2)$ & $E(0_2)$ & $R$\\
\hline
&&&&&&&&\\[-8pt]
SLy4 & $1000$ & 10 & 5.3 & 11.8 & 1.59$\;$ & 1132 & 1911 & 1.68 \\
    & $1250$ & 10.4 & 4.0 & 12.3 & 1.39 & 809 & 1334 & 1.65 \\
SkP & $800$  & 10.5 & 3.7 & 12.6 & 1.45 & 776 & 1306 & 1.68 \\
    & $1000$ & 30.6 & 4.4 & 12.2 & 0.99 & 672 & 1087 & 1.62 \\
DD-PC1 & $728$ & 10.5 & 5.1 & 11.74 & 1.59 & 1092 & 1889 & 1.73 \\
       & $837$ & 9.8  & 4.4 & 11.73 & 1.51 & 925 & 1564 & 1.69 \\
DD-ME2 & $728$ & 10.4 & 4.8 & 11.74 & 1.59 & 1032 & 1794 & 1.74 \\
       & $837$ & 9.9 & 4.2 & 11.73 & 1.50 & 883 & 1499 & 1.70\\
\hline
SU(3)-PDS & & 8.0 & 4.0 & 13.0 & $\sqrt{2}$ & 822 & 1220 & 1.48\\
\hline
\hline
\end{tabular}
\end{center}
\label{Tab1}
\end{table}

\section{SU(3) PDS: an EDF-based approach}
The parameters of the Hamiltonian
$\hat{H}$, Eq.~(\ref{eq:ham}),
derived microscopically from various EDFs,
are given in Table~\ref{tab:para}, along with the parameters
of $\hat{H}_{\rm PDS}$, Eq.~(\ref{h-PDS}),
obtained from a fit to $^{168}$Er~\cite{leviatan1996}.
As discussed in Section~2,
in the latter phenomenological calculation, SU(3)-PDS was
pre-assumed, hence condition~(\ref{PDS-cond}) is satisfied
with $\beta_0\!=\!\sqrt{2}$ and $h_{0}/h_{2}\!=\!2$.
In comparison, in most SCMF calculations, $1.9<h_{0}/h_{2}<2.8$,
consistent with values obtained in global
IBM fits in the rare-earth region~\cite{LevSin99}.
The derived values of $\beta_{0}$ are close
or slightly larger than the SU(3)-PDS value
($\beta_0=\sqrt{2}\approx 1.41$). 
A notable exception are the parameters derived
from the SkP EDF with pairing strength
$V_{0}\!=\!1000$ MeVfm$^{3}$,
which exhibit pronounced large ratio
$h_{0}/h_{2}=6.95$ and small  $\beta_{0}=0.99$.
This is a consequence of the fact that
the corresponding SCMF
energy surface for this case, shown in Fig.~\ref{fig:pes},
is peculiarly soft in the  $\gamma$ deformation, with
a shallow local minimum on
the oblate side. For any chosen EDF,
a larger pairing strength results in a larger
(smaller) value for  $h_{0}/h_{2}$ ($\beta_0$).
\begin{figure*}[t]
\begin{center}
  \includegraphics[width=0.66\linewidth]{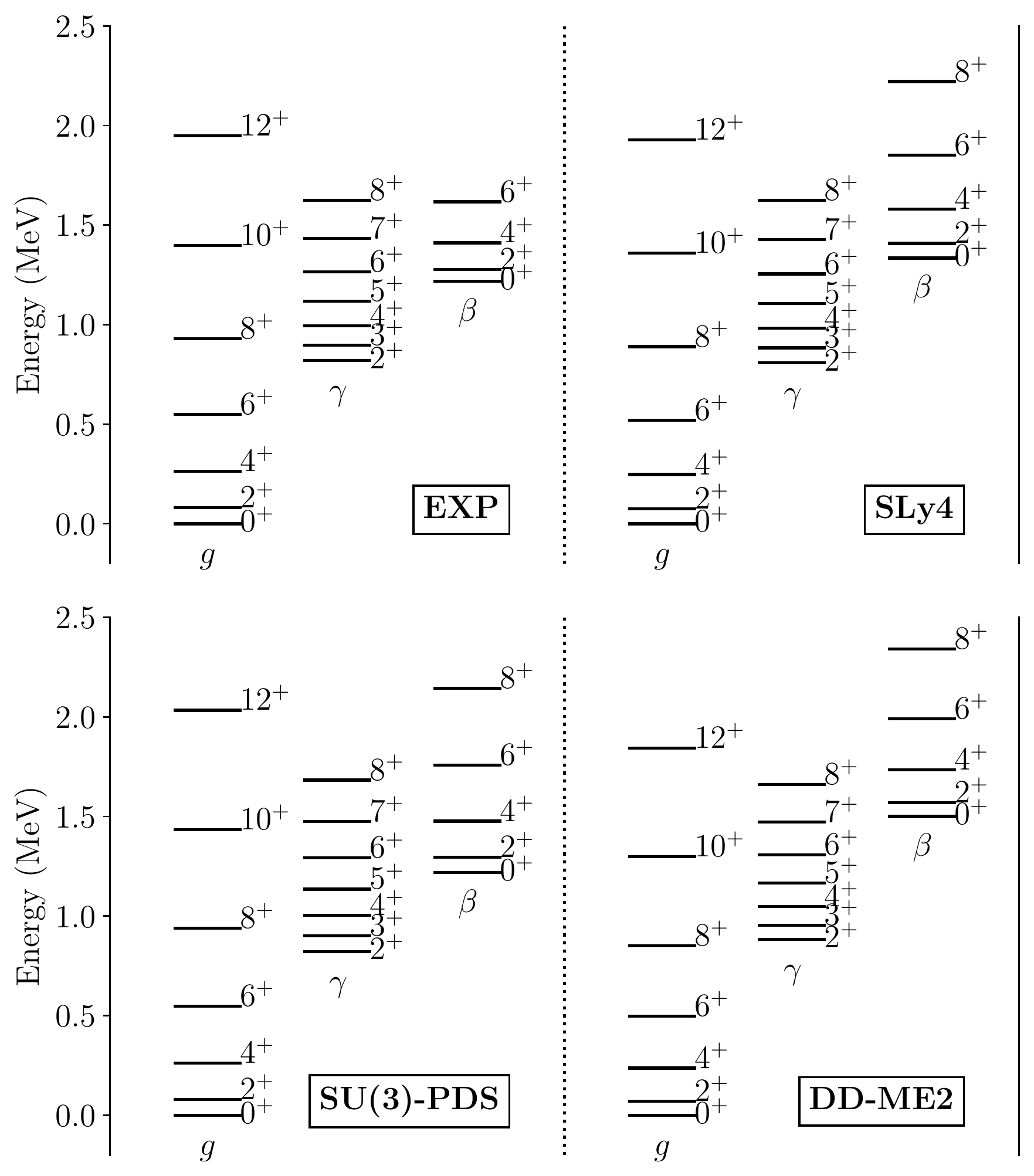}
\caption{\small
Experimental~\cite{nndc} (EXP) and SU(3)-PDS~\cite{leviatan1996}
spectra for $^{168}$Er, compared with the spectra
resulting from EDF-based IBM
calculations for the Skyrme SLy4 EDF with
pairing strength $V_{0}\!=\!1250$ MeVfm$^{3}$, and for
the relativistic EDF DD-ME2 with
$V_{0}\!=\!837$ MeVfm$^{3}$. Adapted from~\cite{NomGavLev21}.}
\label{fig:spectra} 
\end{center}
\end{figure*}
\begin{figure*}[t]
\begin{center}
  \includegraphics[width=0.75\linewidth]{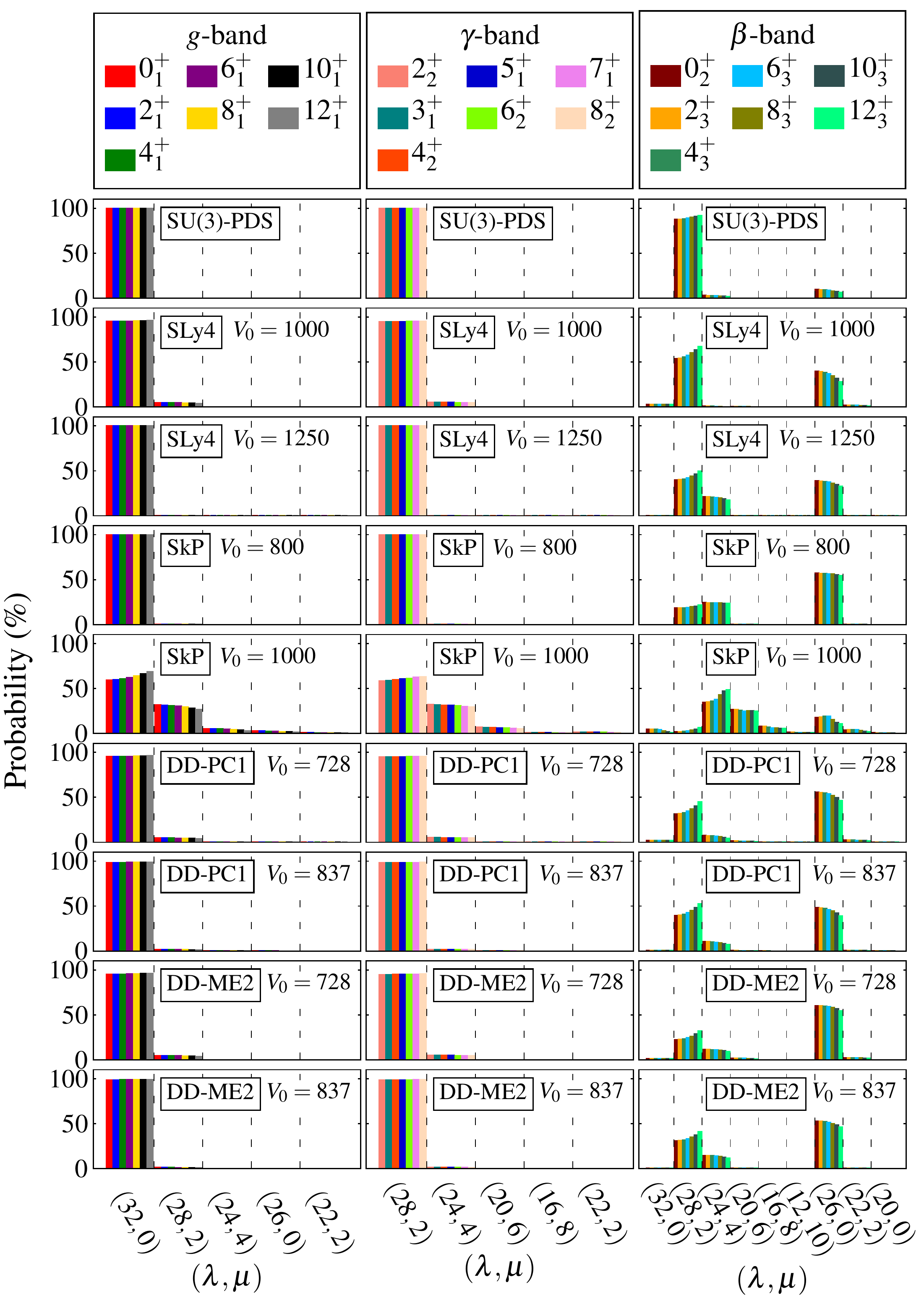}
\caption{SU(3) $(\lambda,\mu)$-decomposition of states in
  the ground ($g$), $\gamma$ and $\beta$ bands,
  for the SU(3)-PDS and various EDF-based calculations.
  Shown are probabilities larger than 0.5~\%.
  The histograms shown from left-to-right for each band,
  correspond to the $L_i$ states listed in the upper
  panels in the order top-to-bottom left-to-right.
Adapted from~\cite{NomGavLev21}.}
\label{fig:decom} 
\end{center}
\end{figure*}

Excitation spectra appropriate for $^{168}$Er are
obtained for each EDF by diagonalizing
the Hamiltonian~(\ref{eq:ham}), using the parameters
in Table~\ref{tab:para} and $N\!=\!16$.
Typical spectra resulting from representative
nonrelativistic and relativistic EDFs
are displayed in Fig.~\ref{fig:spectra}. They
satisfactorily conform with the calculated SU(3)-PDS
spectrum which, in turn, agrees with experimental
spectrum. The bandhead energies, $E(2_2)$ and $E(0_2)$
for the $\gamma$ and $\beta$ bands, and their ratios for
the different cases, are listed in Table~\ref{tab:para}.
In general, the description for the ground and $\gamma$ 
bands is stable with respect to different choices of EDFs.
The description of the $\beta$-band is more case-sensitive
and all EDFs place $E(0_2)$ above the empirical and
SU(3)-PDS values. The following observations are in order.
(i)~The relativistic EDFs generally result in higher
$\beta$-band energies than the Skyrme EDFs.
(ii)~The increase of the pairing strength ($V_0$)
systematically decreases the $\beta$-band energies.
(iii)~The SkP EDF with $V_{0}\!=\!1000$ MeVfm$^{3}$, is
the only case where both $E(2_2)$ and $E(0_2)$ are
placed below the SU(3)-PDS and empirical values.

Analysis of wave functions is a more sensitive measure
to quantify the
similarities and differences in structure between
the EDF-based IBM Hamiltonians and SU(3)-PDS.
Fig.~\ref{fig:decom} shows the SU(3)
$(\lambda,\mu)$-decomposition for member states of the
lowest bands in $^{168}$Er.
For SU(3)-PDS, the ground and $\gamma$ bands are pure
with SU(3) character $(2N,0)$ and $(2N-4,2)$, respectively,
whereas the $\beta$ band contains a mixture
of irreps: $(2N-4,2)$ 87.5~\%, $(2N-6,0)$ 9.6~\%, and
$(2N-8,4)$ 2.9~\%, with $N=16$. 
Remarkably, for all nonrelativistic and relativistic EDFs
considered (except SkP with pairing strength
$V_{0}\!=\!1000$ MeVfm$^{3}$),
the mapped IBM Hamiltonians reproduce very well
the SU(3)-PDS prediction of SU(3)-purity for the ground
and $\gamma$ bands, with probability larger than 95\%.
This clearly demonstrates the robustness of the PDS
notion and its microscopic roots.
The structure of the $\beta$ band is more sensitive to 
the choice of EDF.
Its SU(3) mixing is governed by the values of the
parameters $\beta_{0}$ and ratio $h_{0}/h_{2}$ which,
in turn, reflect the different topology of the
corresponding SCMF surfaces.
Although the dominance of the
$(2N-4,2)$, $(2N-6,0)$, and
$(2N-8,4)$ irreps in the $\beta$ band is generally
observed in all cases, their relative weights differ
from those of SU(3)-PDS.
This may indicate that additional degrees of freedom
not included in the IBM ({\it e.g.}, quasi particles)
contribute to the structure of the $K=0_2$ band in
$^{168}$Er.
Again, the situation is different for the EDF
SkP with  $V_{0}\!=\!1000$ MeVfm$^{3}$ for which
the SU(3) decomposition exhibits large fragmentation.
From all the EDFs considered, the SLy4 and SkP with
$V_{0}\!=\!1250$ and $800$ MeVfm$^{3}$, respectively,
appear to yield spectral properties which are
closest to the SU(3)-PDS predictions for $^{168}$Er
(SU(3) purity for the ground and $\gamma$ bands with
probability 99.8\%).

\section{Conclusions and outlook}
We have shown that
the occurrence of partial dynamical symmetry (PDS) in
nuclei can be justified from a microscopic point of view.
By employing the constrained mean-field methods with
choices of the universal energy density functionals and
pairing interactions, in combination with symmetry
analysis of the wave functions of the mapped IBM
Hamiltonians, we arrived at an efficient procedure to
test and explain the emergence of PDS in nuclei.
An application to $^{168}$Er, has shown that
IBM Hamiltonians derived from known EDFs
in this region, produced eigenstates whose properties
resemble those of SU(3)-PDS.
The fact that these results are valid for
both nonrelativistic and relativistic EDFs
with several choices of pairing strengths,
highlights the robustness of the PDS notion
and its association with properties of
the multi-nucleon dynamics in nuclei.

The results of the present investigation
pave the way for a number of research avenues.
(i)~Exploring the microscopic origin
of other types of PDSs, {\it e.g.}, SO(6)-PDS in
$\gamma$-soft nuclei.
(ii)~When a PDS is found to be manifested empirically in
certain nuclei,
it can be used to constrain,
improve and optimize
({\it e.g.}, choice of the pairing strength)
a given EDF in that region.
(iii)~Exploiting the demonstrated linkage between
the microscopic EDF framework and the algebraic
PDS notion, to predict uncharted regions of exotic
nuclei, awaiting to be explored, where partial symmetries
can play a role.

\section*{Acknowledgements}
The work of A.L. and N.G. is supported by the Israel
Science Foundation. The work of K.N.
is supported by the Tenure Track Pilot Programme of 
the Croatian Science Foundation and the 
\'Ecole Polytechnique F\'ed\'erale de Lausanne,
and the Project TTP-2018-07-3554 Exotic Nuclear Structure
and Dynamics, with funds of the Croatian-Swiss Research
Programme and also
by the QuantiXLie Centre of
Excellence, a project co-financed by the Croatian
Government and
European Union through the European Regional Development
Fund -- the Competitiveness and Cohesion Operational
Programme (Grant KK.01.1.1.01.0004).

\end{document}